\documentclass[%
 reprint,
 superscriptaddress,
 amsmath,amssymb,
 aps,
 pre,
]{revtex4-1}

\usepackage{graphicx}
\DeclareGraphicsExtensions{.pdf}
\usepackage{subfigure}

\graphicspath{{./Fig/}}

\begin{document}

\title{Multilayer weighted social network model}

\author{Yohsuke Murase}
\email{yohsuke.murase@gmail.com}
\affiliation{RIKEN Advanced Institute for Computational Science, 7-1-26, Minatojima-minami-machi, Chuo-ku, Kobe, Hyogo, 650-0047, Japan}
\affiliation{CREST, Japan Science and Technology Agency 4-1-8 Honcho, Kawaguchi, Saitama, 332-0012, Japan}
\author{J\'anos T\"or\"ok}
\affiliation{Department of Theoretical Physics, Budapest University of
Technology and Economics, H-1111 Budapest, Hungary}
\author{Hang-Hyun Jo}
\affiliation{Department of Biomedical Engineering and Computational Science, Aalto University School of Science, P.O. Box 12200, Espoo, Finland}
\affiliation{BK21plus Physics Division and Department of Physics, Pohang University of Science and Technology, Pohang 790-784, Republic of Korea}
\author{Kimmo Kaski}
\affiliation{Department of Biomedical Engineering and Computational Science, Aalto University School of Science, P.O. Box 12200, Espoo, Finland}
\author{J\'anos Kert\'esz}
\affiliation{Center for Network Science, Central European University,
N\'ador u. 9, H-1051 Budapest, Hungary}
\affiliation{Department of Theoretical Physics, Budapest University of
Technology and Economics, H-1111 Budapest, Hungary}
\affiliation{Department of Biomedical Engineering and Computational Science, Aalto University School of Science, P.O. Box 12200, Espoo, Finland}

\begin{abstract}
Recent empirical studies using large-scale data sets have validated the Granovetter hypothesis on the structure of the society in that there are strongly wired communities connected by weak ties. However, as interaction between individuals takes place in diverse contexts, these communities turn out to be overlapping. This implies that the society has a multilayered structure, where the layers represent the different contexts. To model this structure we begin with a single-layer weighted social network (WSN) model showing the Granovetterian structure. We find that when merging such WSN models, a sufficient amount of interlayer correlation is needed to maintain the relationship between topology and link weights, while these correlations destroy the enhancement in the community overlap due to multiple layers. To resolve this, we devise a geographic multilayer WSN model, where the indirect interlayer correlations due to the geographic constraints of individuals enhance the overlaps between the communities and, at the same time, the Granovetterian structure is preserved.
\end{abstract}

\date{\today}
\maketitle

\section{Introduction}

The abundance of data due to the rapid development of the information and communication technology (ICT) has generated entirely new, multidisciplinary approaches in social sciences \cite{lazer2009computational,sen2013sociophysics}, in which physics plays a considerable role both in terms of data analysis and modeling. One of the major challenges in this context is the understanding of the structure of the society, which is crucial for many applications ranging from epidemiology to urban planning. While traditional techniques based mainly on questionnaires focused on small scale organization of the society \cite{wasserman1994social} the new tools enable one to uncover the structure on many scales up to the societal level. A broad range of ICT data has been used to study empirically these questions. Examples include email \cite{kossinets2006empirical}, mobile phone call (MPC) \cite{onnela2007analysis,onnela2007structure,gonzalez2008understanding,pan2011using}, short-message communication, social network services (SNS) \cite{szell2010multirelational}, and scientific collaborations \cite{newman2001structure,menichetti2014weighted}. 

Mobile phone data have a special role in this endeavor as the coverage in the adult population approaches 100\% and much of the interpersonal communication runs today over mobile phones. Therefore the records of the calls can be used to map out the network of social interactions \cite{onnela2007analysis,onnela2007structure,lambiotte2008geographical}. In this mobile call network the famous Granovetter hypothesis about the ``strength of weak ties'' \cite{granovetter1973strength} turned out to be correct. According to this hypothesis links between individuals have different strengths corresponding to the intensity of the relationship, the time spent together, mutual confiding, etc., and the stronger is a tie, the larger is the overlap between the further contacts of those, who form the tie. This local property has severe consequences on the entire structure: The society consists of communities, which are strongly wired and these communities are then connected by weak ties, thus playing an important role to hold society together. The duration or the frequency of calls serves as a natural measure of the strength of ties for mobile phone calls and in this way it was possible to prove the Granovetter hypothesis on this data set \cite{onnela2007structure, onnela2007analysis}. 

In order to demonstrate the global consequences a link percolation analysis was carried out. 
Provided that links are sorted according to their weights, removing the weakest links first one by one results in a sharp transition at a relatively early stage, indicating the fragmentation of the society. In the opposite case, when links were eliminated in the descending order of their weights, the percolation threshold set in at a much higher portion of removed links because strong links are within the communities, where a large number of paths between nodes exist. In this sense, the difference $\Delta f_c$ between the two percolation thresholds can be considered as the measure of the Granovetterian character of the network.

After this empirical verification of the Granovetter hypothesis, the next step was to understand the mechanisms leading to the formation of these structures in a social network by constructing a model, which incorporates basic link-formation processes between individuals. Two main mechanisms were taken into account, namely local and global attachment rules together with tie strength reinforcement \cite{kumpula2007emergence,jo2011emergence}. Here the local and global attachment rule correspond to cyclic and focal closure mechanisms \cite{kossinets2006empirical}, with the former referring to the link formation with one's network neighbors, or with friends of friends and the latter to the attribute-related link formation which is independent of the local network topology. The reinforcement step corresponds to the general observation that social ties get strengthened by using them. With these simple processes, the complex Granovetterian weight-topology relation of social networks could be successfully reproduced as demonstrated for the large value of $\Delta f_c$ \cite{kumpula2007emergence}.

The community structure of complex networks is an extensively studied topic \cite{fortunato2010community}. The identification of communities or structural modules, i.e., groups of nodes having more connections among themselves than outside the group is a highly nontrivial task and much effort has been devoted to its solution (see, e.g., \cite{newman2006modularity, blondel2008fast,rosvall2008maps}.) Most of the methods produce a partition of the network, meaning that a node can belong to only one community. However, as pointed out in \cite{palla2007quantifying} this cannot lead to an appropriate description of many complex networks, especially of social ones, where there is usually considerable overlap between the communities due to the fact that nodes can belong simultaneously to several of them. A number of algorithms have been suggested to uncover overlapping communities \cite{palla2007quantifying, lancichinetti2009detecting, ahn2010link}. 

The community detection method of Ahn \textit{et al.} \cite{ahn2010link} was based on the identification of link communities; at the same time they suggested a remarkable mechanism as the origin of overlapping communities. Using the language of social networks (what we are interested in here), they propose that a person can be in different types of relationships, like kinship, collaboration, friendship, etc. Moreover, people are switching their social contexts and communication channels depending on the occasions, and the social network should strongly depend on the context \cite{jo2012spatiotemporal,jo2013contextual}. To handle these aspects, it is necessary to represent the social networks as a multilayer or multiplex network \cite{Kivela_JCN2014,boccaletti2014structure,jo2006immunization}, where each layer corresponds to a different type of relationship. Since these contexts are hardly distinguishable from the available data, the networks observed in this way are usually considered as projections or an aggregate of multiple layers. Such a projection of multilayer networks should be in line with the observed stylized facts faithfully with empirical data, such as Granovetter-type structure. An important aspect of multilayer structure is missing both from the original Granovetter paper \cite{granovetter1973strength} and the above described model \cite{kumpula2007emergence}. The aim of the present paper is to investigate the possibilities to model the combination of the multilayer structure of the society with the Granovetterian relationship between tie strengths and topology. In order to do so, we start from the simple, single-layer model by Kumpula \textit{et al.} \cite{kumpula2007emergence} and introduce the multilayer structure in different ways.

This paper is organized as follows. In the next section, a naive multilayer network is investigated and it is shown that it leads to a break down of the Granovetter-type structure as correlations are suppressed. We therefore introduce the copy-and-shuffle model, where a parameter tunes the correlations. We find a regime with $\Delta f_c$ significantly different from zero, however, there the average number of overlapping communities, a node participates in, is low. To overcome this difficulty, we formulate a model in Sec.~\ref{sec:geographic_multilayer}, where correlations are caused by the dependence on the geographic distance. This model has a parameter region where both Granovetterian structure and a considerable enhancement of the average overlap are observed. The last Section is devoted to a summary and discussion.

\section{Multilayer Weighted Social Network (WSN) Model}

\subsection{Single-layer WSN Model}

Let us first summarize the original WSN model by Kumpula \textit{et al.} \cite{kumpula2007emergence}. It considers an undirected weighted network of $N$ nodes. The links in the networks are updated by the following three rules. The first rule is called {\it local attachment} (LA). Node $i$ chooses one of its neighbors $j$ with probability proportional to $w_{ij}$, which stands for the weight of the link between nodes $i$ and $j$.

Then, node $j$ chooses one of its
neighbors except
$i$, say $k$, randomly with probability proportional to $w_{jk}$. If node $i$ and $k$ are not connected, they are connected with probability $p_{\Delta}$ with a link of weight $w_{0}$, but if they are already connected this link weight and the other two link weights $w_{ij}$ and $w_{jk}$ in a triangle are increased by $\delta$.
The second rule is {\it global attachment} (GA), where if a node has no links or otherwise with probability $p_r$, it is connected to a randomly chosen node with weight $w_{0}$.
Finally, the third rule {\it node deletion} (ND) is introduced to the model, where with probability $p_{d}$, a node loses all its links.
At each time step, LA, GA, and ND are applied to all nodes. Starting from a network without any links, the network reaches a statistically stationary state after a sufficient number of updates. As a function of the reinforcement parameter $\delta$ this model shows a gradual transition from a module free topology to a Granovetterian structure with strongly wired communities connected by weak ties. 

\subsection{Generalization to the multilayer case}\label{subsec:model_multi_layer}

In order to study multilayer effects we generalize the single-layer WSN model in the following naive way. We consider $L$ layers of the same set of nodes and we assume that each layer corresponds to a different type of relationship or communication context. For each layer, we independently construct a network in the same way as in the original single-layer WSN model. For simplicity, the same parameters are used for all the layers. After the stationary networks are constructed in each layer, the aggregate network is constructed by summing up the edge weights: $w_{ij}=\sum_{k=1}^L w_{ij}^k$, where $w_{ij}^k$ is the weight of the link between nodes $i$ and $j$ in the $k$-th layer \footnote{Significant difference was not found even if we take the maximum of the link weights instead of the sum. This is because majority of the links belong to one of the layers.}. It is this aggregate network for which we expect the Granovetterian structure.

\begin{figure}
\begin{center}
\includegraphics[width=.45\textwidth]{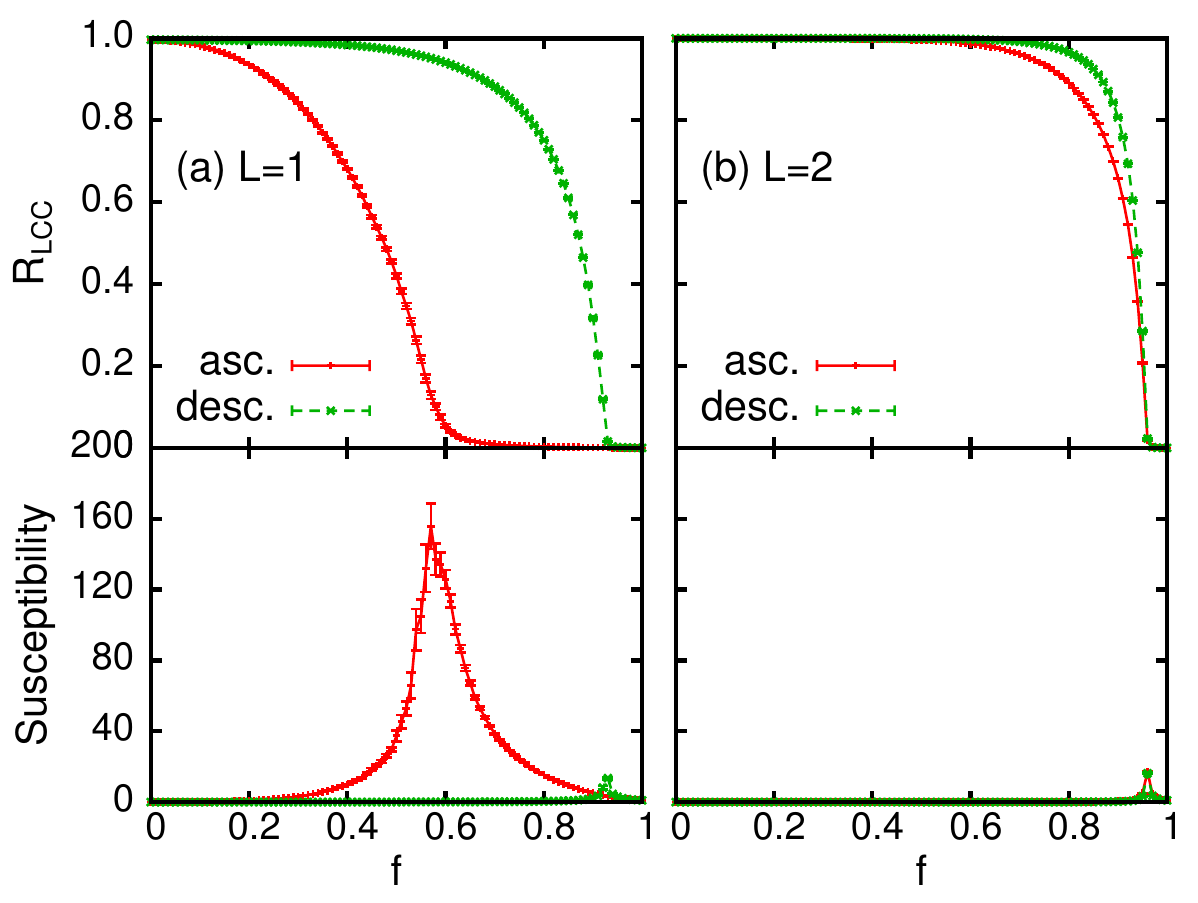}
\caption{(Color online) Link percolation analysis for $L=1$ (left) and $L=2$ (right). The upper figures show the relative size of the largest connected component, $R_{LCC}$, as a function of the fraction of the removed links $f$. The lower figures show the susceptibility $\chi$. Red solid (green dashed) lines correspond to the case when links are removed in ascending (descending) order of the link weights. The error bars show standard errors.
}
\label{fig:percolation_L1L2}
\end{center}
\end{figure}

\begin{figure}
\begin{center}
\includegraphics[width=.45\textwidth]{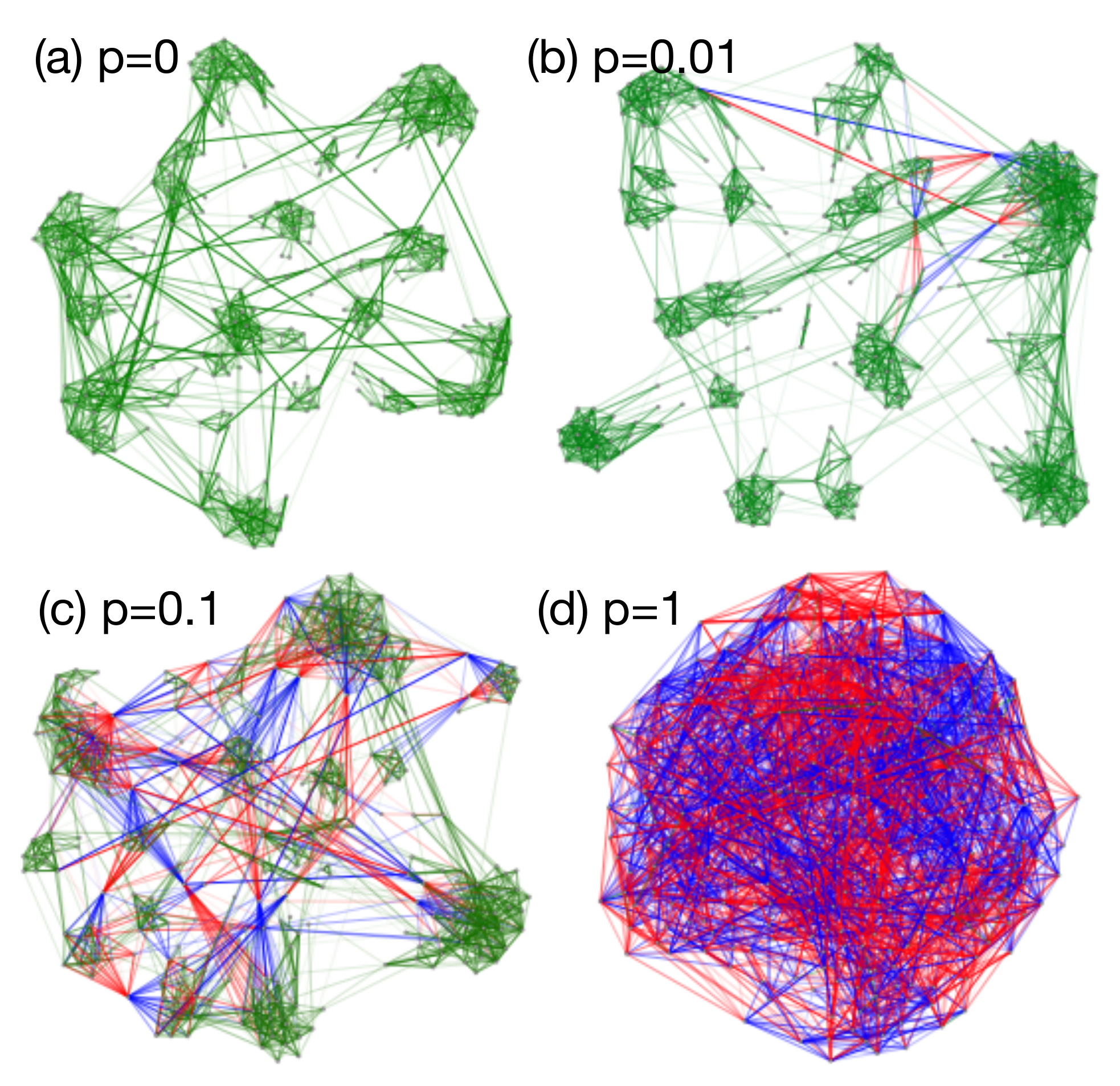}
\caption{
(Color)
Snapshots of the copy-and-shuffle model with different $p$ shuffling parameter values and $N=300$. Red (blue) links are in the first (second) layer, and green links are in both layers. 
}
\label{fig:snapshot_multilayer}
\end{center}
\end{figure}

\begin{figure}
\begin{center}
  \includegraphics[width=.5\textwidth]{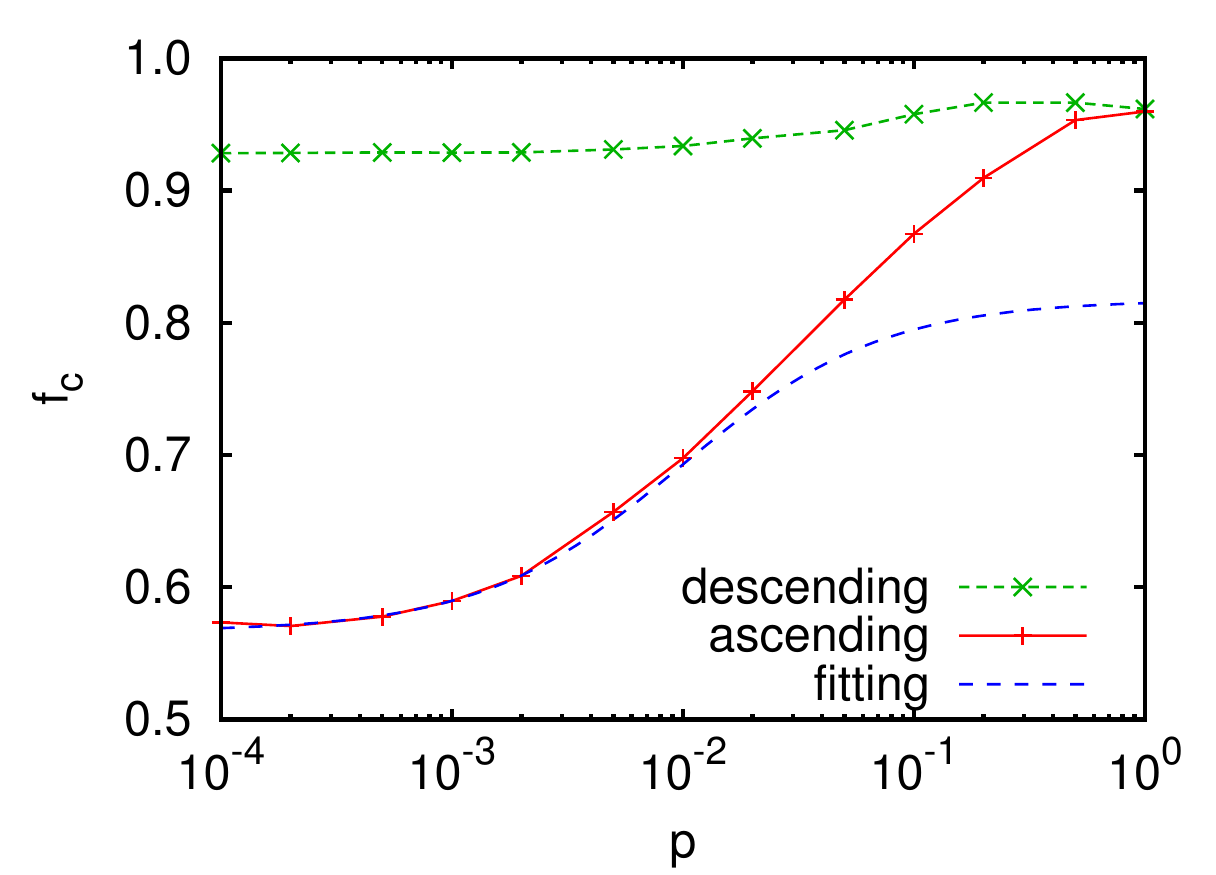}
\caption{
  (Color online)
Percolation thresholds for various shuffle fraction values $p$ for the copy-and-shuffle model. The green upper and red lower lines denote the critical points $f_c^d$ and $f_c^a$, respectively. The critical points are determined by the peak of the susceptibility. The points are calculated for $50$ independent runs. The blue dashed line is calculated using Eq.~(\ref{eq:copyshufflemodel}).
}
\label{fig:f_c}
\end{center}
\end{figure}

In the following, $N = 50000$, $p_r = 0.0005$, $p_{\Delta} = 0.05$, $p_d = 0.001$, $\delta=1$, and $w_0 = 1$ are used. The results are obtained after $25 \times 10^3$ time steps and averaged over $50$ realizations. To see whether the multilayer model reproduces a realistic social network of the kind the mobile phone call (MPC) graph is a proxy \cite{onnela2007analysis,onnela2007structure}, a link percolation analysis is carried out for the model. We removed fraction $f$ of the links from the generated networks in both ascending and descending orders, and measured the relative size $R_{LCC}$ of the largest connected component and the normalized susceptibility $\chi = \sum n_{s} s^2 / N$, where $n_s$ is the number of components of size $s$ and the sum is taken over all but the largest component. At the percolation threshold the order parameter $R_{LCC}$ vanishes and $\chi$ diverges in the thermodynamic limit. For finite systems the former quantity shows a fast decay and the latter one a sharp peak at the threshold value $f_c$. The significant difference $\Delta f_c$ in the thresholds for the two sequences of link removal is characteristic by the Granovetter structure; $\Delta f_c = f_c^d-f_c^a$, where the upper index $d$ ($a$) stands for descending (ascending) sequences of removed links.  

Figure~\ref{fig:percolation_L1L2} shows $R_{LCC}$ and $\chi$ as a function of $f$ for a single-layer network ($L=1$) and a double-layer network ($L=2$).
The two plots in each figure show the results for ascending and descending orders. For $L=1$ we get $\Delta f_c \approx 0.35$,
while for $L=2$ the figure shows that the percolation threshold for ascending order $f_c^a$ is not significantly different from that for descending order $f_c^d$ (i.e., $\Delta f_c \approx 0$).

The percolation thresholds for $L=2$ are approximately the same, $f_c \approx 0.95$, indicating that the introduction of a second layer destroys the Granovetterian structure. The percolation threshold agrees well with that of an Erd\H os-R\'enyi (ER) random network having the same average degree $\langle k \rangle$ as the simulated model: $f_c = 1 - 1/\langle k \rangle$ with the measured $\langle k \rangle = 21.9$. (Note that this is twice the average degree of a single layer.) This observation shows that combining already two independent layers from the original single-layer WSN model leads to a high level of randomization in the aggregate \footnote{Of course, the resulting network has a topology different from an ER random graph as it has high clustering by construction.}. One may think that the observed effect is due to the increasing total degree when two layers are merged.
However, we carried out simulations, where 
the total degree was controlled by $p_{\Delta}$ and found that for $L=2$ the thresholds are always very close to each other; $\Delta f_c \approx 0$.

\subsection{Copy-and-shuffle WSN model}

Due to the fact that merging two layers of WSN models destroys the Granovetterian structure, we investigated how the correlation between layers affects the properties of the network.
We created the second layer by copying the first layer and then shuffled the fraction $p$ of the nodes in the second layer. Shuffling nodes $i$ and $j$ means that all original links $(i,k)$ become $(j,k)$ and vice versa. This is just a relabeling of the nodes in that layer, meaning that the topology remains the same, i.e., both layers correspond to single-layer WSN models but with increasing $p$ the correlations between them decrease. This is called the ``copy-and-shuffle'' model (see Fig. \ref{fig:snapshot_multilayer}).

When $p=0$, the aggregate network is equivalent to the single-layer
network whose link weights are doubled. For $p=1$, it is the same as
the double-layer model; the Granovetterian structure gets entirely
destroyed by randomization. By controlling $p$ between $0$ and $1$, a
transient behavior is observed. Figure \ref{fig:f_c} shows how the threshold values $f_c^a$ get closer to $f_c^d$ as $p$ is increased and for $p \to 1$ we get $\Delta f_c \to 0$. The reason is that strong links in the second layer connect the communities more randomly since the correlation between the first and the second layer diminishes.

The randomization has the consequence that the percolation threshold $f_c^a$ gets closer to that of the corresponding Erd\H os-R\'enyi random network. However, for a reasonably large range of $p$, we can clearly differentiate $f_c^a$ and $f_c^d$ thus the similarity between the layer assures the Granovetterian structure. 

The gradual transition can be understood in the following way. Let us make the assumptions that the original network (first layer) is composed of strongly connected groups interconnected by weak links [see Fig.~\ref{fig:snapshot_multilayer}(a)] and the average size of these groups is $\overline{s}$ which is small and independent of the total number of agents $N$. Eventually the number of groups is $N_g=N/\overline{s}$. In the link percolation analysis starting from the weak links (ascending order) we can consider the groups as ``supernodes'' and we have to solve the percolation problem for the links connecting them.

Not only the intergroup links turn out to be weak but also some intragroup ones. Let us denote the number of links by $M$, the total number of weak links by $M_w$ and the intergroup weak links by $M_g$ which are a subset of $M_w$. Let us remove the $f^a$ fraction of the total links in ascending order. In this case the number of weak links gets $M_w(f^a)=M_w-Mf^a$ since we only removed weak links. Thus it is clear that this approximation will not work for $f^a>M_w/M$. The intergroup links are removed by the same rate as weak links, so the number of intergroup links after removing the $f^a$ part of the total links in ascending order is
\begin{equation}\label{eq:intergrouplinknumber}
M_g(f^a)=M_g\left(1-f^a\frac{M}{M_w}\right).
\end{equation}

In order to have quantitative results we need the number of strong links. This can be estimated if we assume that $M_w=M(\langle k\rangle-2)/\langle k\rangle$. This means that each node has two strong links while the rest are weak. This can be justified
by considering the effect of cyclic closure: The cyclic closure is the most frequent interaction which strengthens two links of a focal node with positive feedback of their weights.
Assuming a random network for the groups at the percolation threshold, 
one should get $M_g(f^a_c)=N_g/2$. This gives $\langle k_g\rangle\equiv 2M_g/N_g=3.26$ for the single-layer model, where $\langle k_g\rangle$ is the average degree of the connections between the groups, i.e., that of the supernodes. Snapshots of the single-layer model as in Fig. \ref{fig:snapshot_multilayer}(a) justify the low contact number for the groups.

The shuffling of agents creates new connections from one group to another.
These connections increase the intergroup connection degree by
\begin{equation}
\langle k_g\rangle(p) \approx \langle k_g\rangle(p=0)+p\overline{s}\langle
k\rangle.
\end{equation}
Thus there is a linear increase of the average degree of the groups with $p$.

Now we can use Eq.~(\ref{eq:intergrouplinknumber}) to go in the reverse direction, namely that knowing $M_g/N_g$ we can get the percolation threshold (note that $M_w$ and $M$ also depend on $p$ in a trivial way):
\begin{eqnarray}\label{eq:copyshufflemodel}
  \nonumber
f_c^a&=&\frac{M_w(p)}{M(p)}\left(1-\frac{N_g}{2(M_g+pN\langle k\rangle)}\right)\\
&=&\frac{\langle k\rangle-2}{\langle k\rangle} \left(1-\frac{1}{ \langle k_g\rangle(p=0)+2\overline{s}\langle k\rangle p}\right).
\end{eqnarray}

The average size of the groups for the single layer can be obtained by an infomap analysis \cite{rosvall2008maps} and was found to be $\overline{s}=15.1$. The resulting curve is shown in Fig.~\ref{fig:f_c} as a dashed line and is compared to the empirical threshold values. The calculated line fits the initial part very well, where the above picture is expected to work.

\begin{figure}
\begin{center}
\includegraphics[width=.48\textwidth]{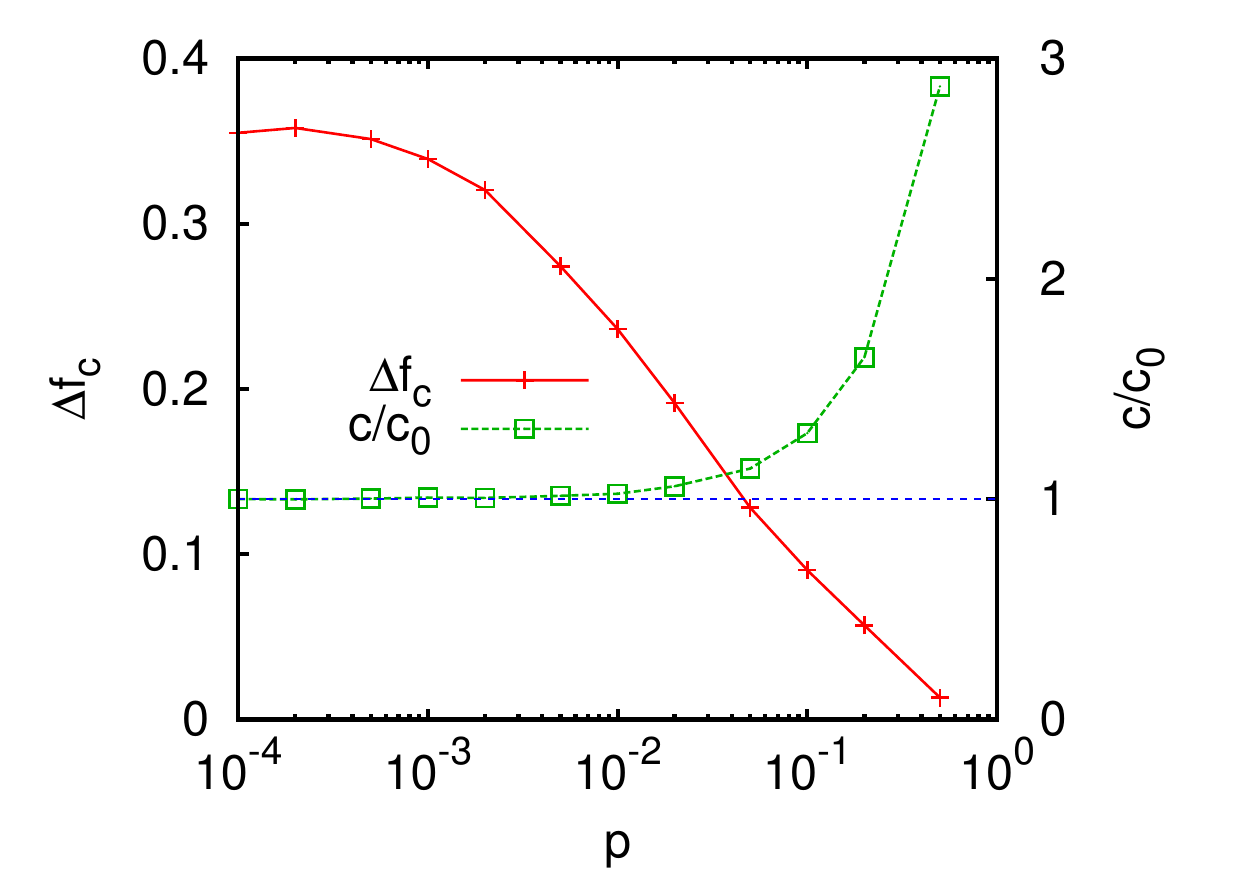}
\caption{
(Color online)
Characteristic quantities for the copy-and-shuffle multilayer WSN model. The difference $\Delta f_c$ between the percolation thresholds decreases with the shuffling probability $p$, while $\overline{c}/\overline{c}_0 $, the ratio of the average number of communities a node belongs to at parameter value $p$ and $p=0$, decreases. There is no regime, where $\Delta f_c$ is significantly larger than zero and $\overline{c}/\overline{c}_0$ is considerably larger than one. The results are averaged by $50$ independent samples and the errors are smaller than the symbol size.
}
\label{fig:df_cc_cs}
\end{center}
\end{figure}

The copy-and-shuffle model produces a region of $p$, where a multilayer Granovetterian structure exists. Now we have to check whether our construction has lead to enhancement of the overlapping of the communities, too. We have analyzed the aggregate networks by the method of Ahn \textit{et al.} \cite{ahn2010link} and calculated $\overline{c}/\overline{c}_0 $, the ratio of the average numbers of communities a node belongs to at parameter value $p$ and $p=0$ \footnote{We get even for $p=0$ a value larger than 1 ($\overline{c}_0=2.96$. This is due to the fact that the method of \cite{ahn2010link} is a partition of the link graph producing a large number of single links (diads) as communities, which enhance the overlap.}. We expect that $\overline{c}/\overline{c}_0 $ should increase as shuffling goes on. Figure \ref{fig:df_cc_cs} shows the dependence of this quantity on $p$.
The overlap starts to increase only when the Granovetterian correlation between link weight and topology is already wiped away. In the next section we make another attempt to produce a model, where the Granovetterian structure and overlapping communities coexist.

\section{Geographic multilayer WSN model}\label{sec:geographic_multilayer}

The above results show that some correlations are needed between layers in order to have $\Delta f_c$ significantly different from zero for a multilayer model.
Previous studies have reported that there are strong geographic constraints on social network groups even in the era of the Internet \cite{onnela2011geographic}
and this is reflected in the MPC data \cite{krings2009urban,lambiotte2008geographical, expert2011uncovering}. For example, intercity communication intensity is inversely proportional to the square of their Euclidean distance, which is reminiscent of the gravity law \cite{krings2009urban,lambiotte2008geographical}. 

Motivated by these observations, we consider now a model embedded into a two-dimensional geographic space. At the beginning of the simulation nodes are distributed randomly in the unit square with periodic boundary condition. These geographic positions are fixed and shared by all the layers. We assume that the probability for making a new connection in the global attachment (GA) step in the WSN model is higher if the two nodes are geographically close. The probability that node $i$ makes a new connection to node $j$ by GA is proportional to $r_{ij}^{-\alpha}$, where $r_{ij}$ is a distance between nodes $i$ and $j$, and $\alpha$ is a new parameter controlling the dependence on geographic distance as in \cite{kosmidis2008structural,daqing2011dimension}. When $\alpha = 0$, this probability is independent of the geographic distance, thus the model is equivalent to the uncorrelated multilayer model we presented in the previous section. When $\alpha$ is larger, the nodes tend to be connected with geographically closer nodes yielding the correlation between the networks in different layers. Since only nonconnected pairs are considered, the probability for node $i$ to make connection with a node $j$ which is not yet connected to $i$ is given by
\begin{equation}\label{eq:alphadef}
p_{ij} = \frac{ r_{ij}^{-\alpha} }{\sum_{k \in S_i} r_{ik}^{-\alpha} },
\end{equation}
where $S_i$ is the set of the nodes not connected to the node $i$. The other rules such as LA or ND are kept the same as in the original WSN model.

\begin{figure}
\begin{center}
\includegraphics[width=.45\textwidth]{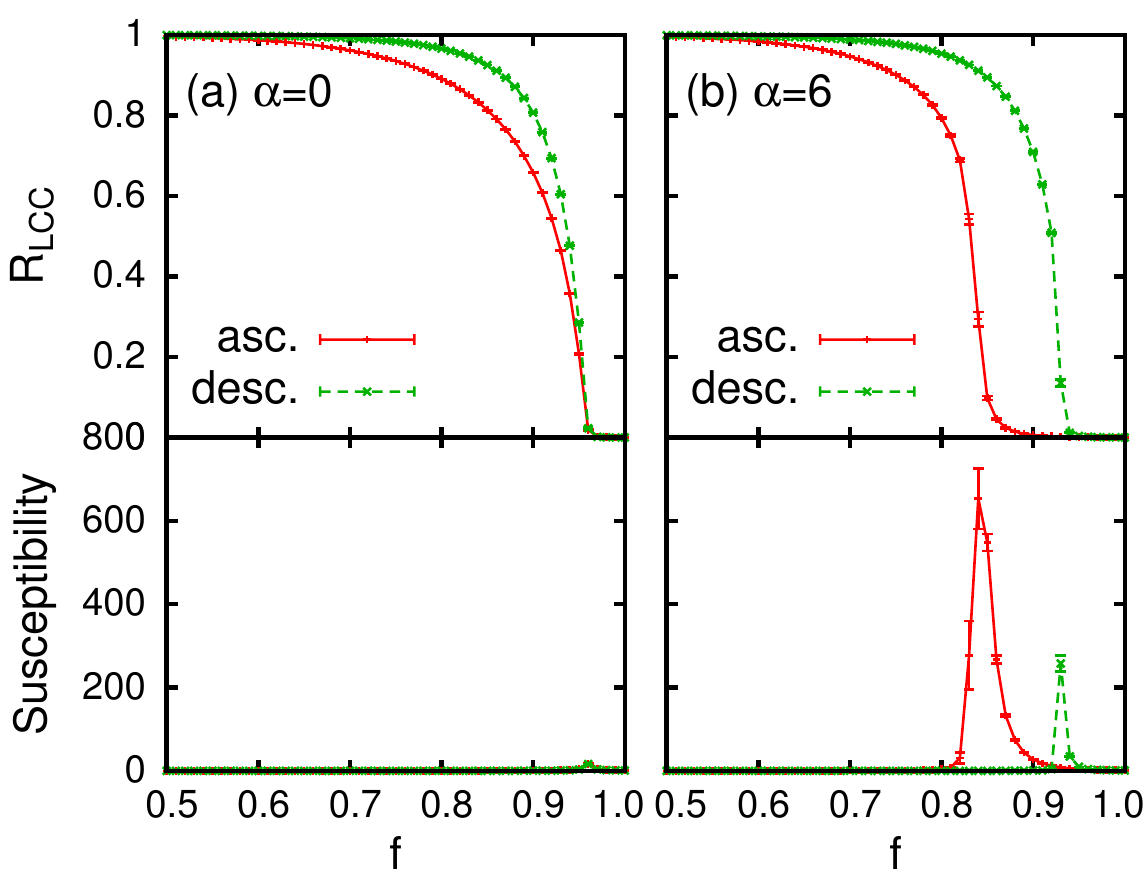}
\caption{
(Color online)
Link percolation analysis for (a) $\alpha=0$ and (b) $\alpha=6$. The upper figures show the relative size of the largest connected component, $R_{LCC}$, as a function of the fraction of the removed links $f$. The lower figures show the susceptibility $\chi$. Note that the scale of the horizontal axis is different from Fig.~\ref{fig:percolation_L1L2}. Red solid (green dashed) lines correspond to the case when links are removed in ascending (descending) order of the link weight. The results are obtained by $50$ independent samples. The error bars show standard errors.
}
\label{fig:percolation_L1L2_geographic}
\end{center}
\end{figure}

\begin{figure}
\begin{center}
\includegraphics[width=.48\textwidth]{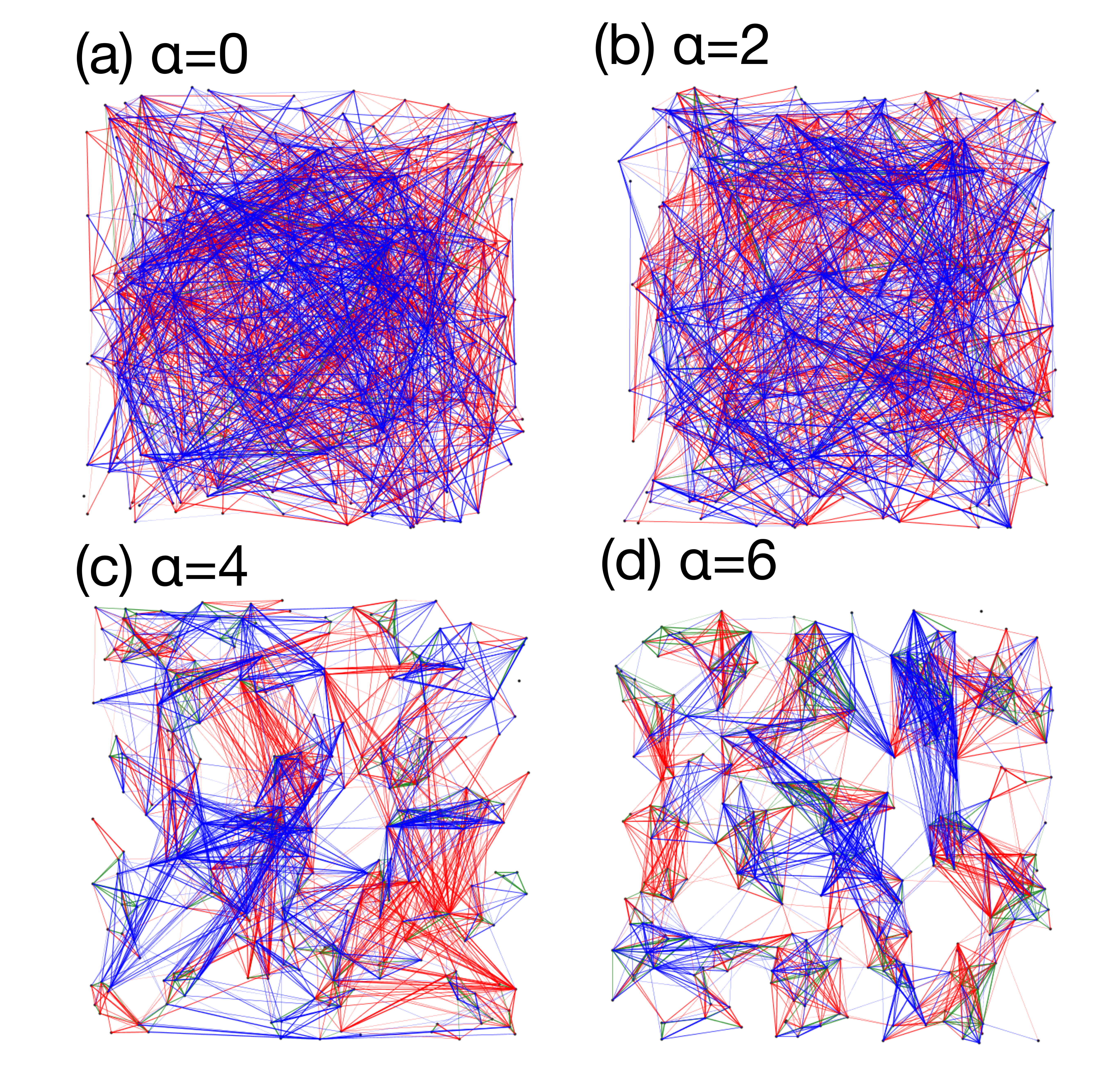}
\caption{
(Color)
Sample of double-layer networks of the geographic model for $N=300$. Links only in the first and the second layers are shown in blue and red, respectively, and links shared by both layers are depicted in green.}
\label{fig:snowball_samples_geographic}
\end{center}
\end{figure}

Figure \ref{fig:percolation_L1L2_geographic} shows the results for link percolation analysis for the geographic model with $\alpha=0$ and $6$. Because the network for larger $\alpha$ has a smaller average degree, we used a larger value of $p_r$ ($0.002$) in order to keep the average degree comparable to the results for the nongeographic model ($\langle k \rangle =18.0$ for $\alpha=6$ and $\langle k \rangle =27.6$ for $\alpha=0$). As shown in the figure, the network for $\alpha=6$ exhibits a Granovetterian structure as $f_c^a$ and $f_c^d$ are significantly different with $\Delta f_c \approx 0.1$.

\begin{figure}
\begin{center}
\includegraphics[width=.46\textwidth]{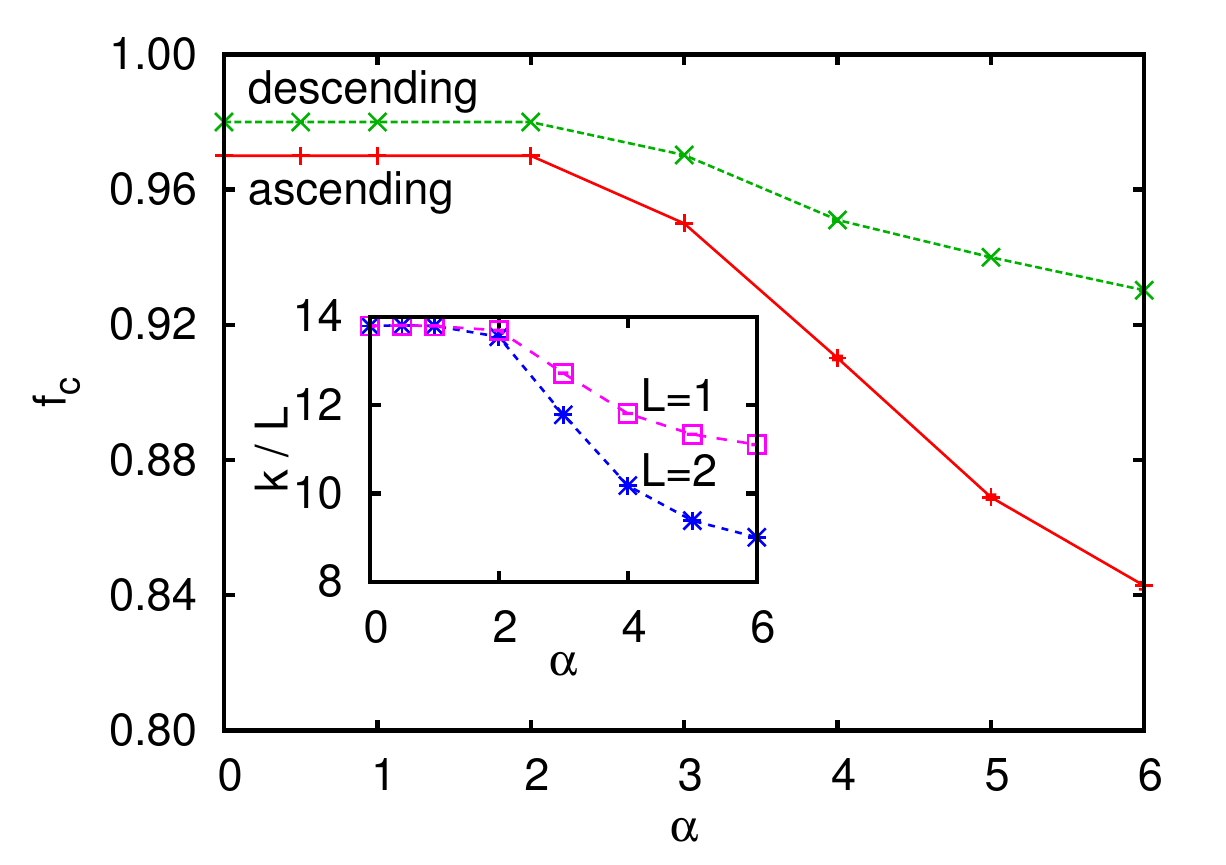}
\caption{
(Color online)
Percolation thresholds $f_c^a$ and $f_c^d$ for the geographic model as a function of $\alpha$. The percolation thresholds are determined as the point where maximum susceptibility is observed and then averaged over $50$ independent samples. (Inset) Average degrees divided by $L$ as a function of $\alpha$ for $L=1$ and $L=2$. The values for $L=2$ are smaller than those for $L=1$ when $\alpha$ is sufficiently large because there are links appearing in both layers.
}
\label{fig:f_c_alpha}
\end{center}
\end{figure}

Small samples of networks ($N{=}300$) for different $\alpha$ are shown in Fig.~\ref{fig:snowball_samples_geographic}. While the network for $\alpha=0$ and $2$ look similar to the uncorrelated double-layer nongeographic network in Fig.~\ref{fig:snapshot_multilayer}(d), the networks for larger $\alpha$ clearly show a nice community structure. About $19$ percent of the links are shared by two layers for $\alpha=6$, while less than $0.1$ percent of the total links are shared for $\alpha=0$; see also the inset of Fig.~\ref{fig:f_c_alpha}. This already indicates the possibility of overlapping communities.

\begin{figure}
\begin{center}
\includegraphics[width=.5\textwidth]{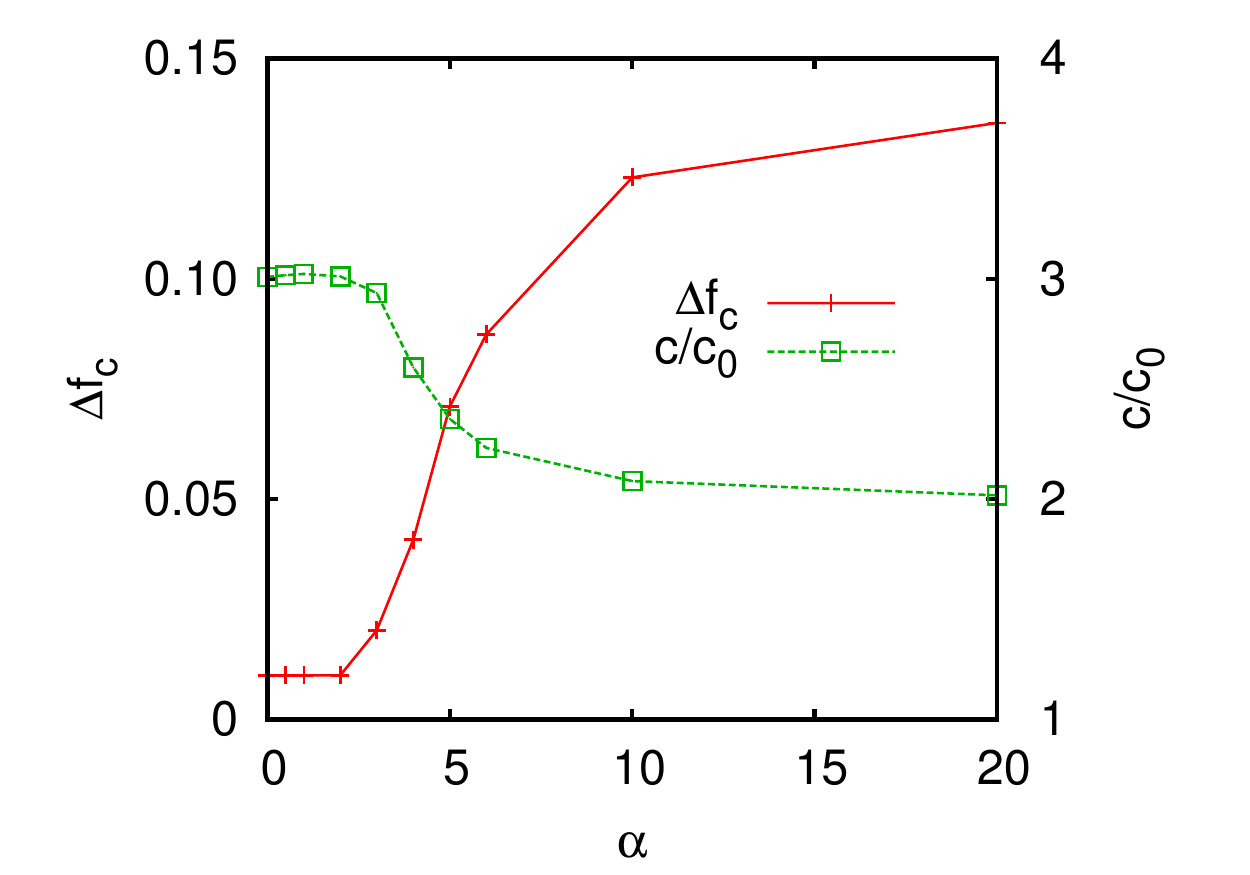}
\caption{
(Color online)
This figure is a similar plot in Fig. \ref{fig:df_cc_cs} for the geographic multilayer WSN model. Here $\Delta f_c$ and $\overline{c}/\overline{c} _0$ are shown as a function of $\alpha$. Note that for this model $\overline{c}_0$ also depends on $\alpha$. For $\alpha\ge 6$ we have $\Delta f_c$ significantly larger than 0 and $\overline{c}/\overline{c} _0$ close to 2.
The results are obtained by $50$ independent samples.
}
\label{fig:df_cc_geo}
\end{center}
\end{figure}

The dependence of $f_c$ on $\alpha$ is shown in Fig.~\ref{fig:df_cc_geo}, which summarizes the main results for the geographic model. $\Delta f_c$ becomes larger with increasing $\alpha$ and seems to get saturated around 0.15. The ratio $\overline{c}/\overline{c}_0$ decreases rather rapidly and reaches the limit value of 2. This means that for sufficiently large $\alpha$ we have {\it both} Granovetterian properties and the enhancement of the number of overlapping communities due to the multilayer structure. We note that for $\alpha>4$ both the average degree in one layer $\langle k(L=1)\rangle \simeq11$ and $\overline{c}_0(\alpha>4)\simeq 3$ is the same as in the nongeographic case indicating similar structure.


We note that neither the percolation thresholds nor the average degree show
significant dependence on $\alpha$ for $\alpha < 2$. We speculate that
this is because the network dimension becomes infinite for $\alpha <
2$ even when it is embedded in a two-dimensional space \cite{daqing2011dimension}.
Since the dimensionality of the network is finite, the clustering
coefficient for the network is higher compared to the network with
smaller $\alpha$. (For $\alpha=0$, $2$ and $6$, clustering
coefficients are $0.23$, $0.24$, and $0.55$, respectively.) This also
explains the change in the average degree. If a link and its
neighboring link are selected by LA, the probability that the third
link closing the triangle is already there will be higher for higher
$\alpha$ thus the number of links newly created by LA is smaller
leading to the decrease in the average degree.

\section{Summary and Discussion}

Our aim in this paper has been to model two important properties of the social network: Its Granovetterian structure and the large amount of overlapping communities due to its multilayer character. We introduced the difference $\Delta f_c$ of the percolation thresholds $f_c^d$ and $f_c^a$ as a single variable characterizing
the 
weight-topology relation and $\overline{c}/\overline{c}_0$, the ratio of the average number of communities a node participates in for the multilayer and the single-layer networks. We expect from a model satisfying our goal simultaneously a $\Delta f_c$ considerably different from zero and $\overline{c}/\overline{c}_0$ significantly larger than one. 

The naive introduction of multiple layers of single-layer WSN models breaks the Granovetter-type weight-topology relation since the communities in one layer get connected by strong ties in another layer. If we control the amount of randomness by the shuffling probability $p$ and start from replicas of single-layer models then we arrive at a multilayer model, which, however, has no region of the control parameter $p$, where both required properties can simultaneously be observed.

In order to maintain both requirements, we introduced an extension of the single-layer model such that each node has a geographic position and that geographically close nodes have more chance to form a link via GA. The multilayer model consists then of a combination of such single layers. Controlling the distance dependence by the exponent $\alpha$ [as defined in Eq. (\ref{eq:alphadef})] we conclude that for $\alpha \ge 6$ we have a multilayer weighted social network, which has both the Granovetterian structure and the enhanced community overlap.

We note here that previous studies on MPC data have revealed that intercity communication density is inversely proportional to the square of the distance \cite{krings2009urban,lambiotte2008geographical}. Regarding the geographic model, the relation between $\alpha$ and the exponent characterizing intercity communication density is not trivial because the links created by LA are not affected by the geographic position. Furthermore, we assumed that the position of the nodes are uniformly distributed, which is clearly an idealized aspect of the model since we know that population usually aggregates around city areas.

Our results have several implications. First, they show that geographic correlations play a key role to change the picture drastically in a multilayer weighted network similarly to what was observed for interdependent networks \cite{parshani2010inter}. Moreover, although the models we studied are strong simplifications of the society, we believe that they have their role in the investigation of social structures. In particular, such models enable one to study the special effects of the Granovetterian and the overlapping community structure on dynamic phenomena like spreading. 

Communities organize themselves along common attributes like sharing working places, classes at universities, joint interest, e.g., in sport, residential districts etc. \cite{kosmidis2008structural}. Geographic proximity is just one of the possibilities and other attributes can play an important role in the formation of network as well.
Future work is needed to find out how to treat explicitly these attributes and their impact on the formation of the network.

Our models have also implications for further empirical studies.
Unfortunately, most datasets contain only information about one
channel of communication, which substantially restricts sampling of
relationships even in the case of mobile call networks. An alternative
approach is ``reality mining'', where a limited number of volunteers (of
the order of one hundred) carry devices, which record several channels of
communication, including face-to-face encounters \cite{eagle2006reality}. This methodology could
pave the way for studies of the effects of the multilayer character
of human society, especially from the points of view presented in this
paper. 

\begin{acknowledgments}
Y.~M. appreciates hospitality at Aalto University and H.-H.~J. acknowledges financial support from the Aalto University postdoctoral program. The systematic simulations in this study were assisted by OACIS \cite{murase2014tool}.
J.~T. acknowledges financial support from the European Union and the European Social Fund through project FuturICT.hu (Grant No.: TAMOP-4.2.2.C-11/1/KONV-2012-0013). J.~K. acknowledges support from EU Grant No. FP7 317532 (MULTIPLEX).
\end{acknowledgments}

\appendix
\section{Trial with other parameters}\label{section:try_other_parameters}
\begin{figure}
\begin{center}
\subfigure{
\includegraphics[width=.47\textwidth]{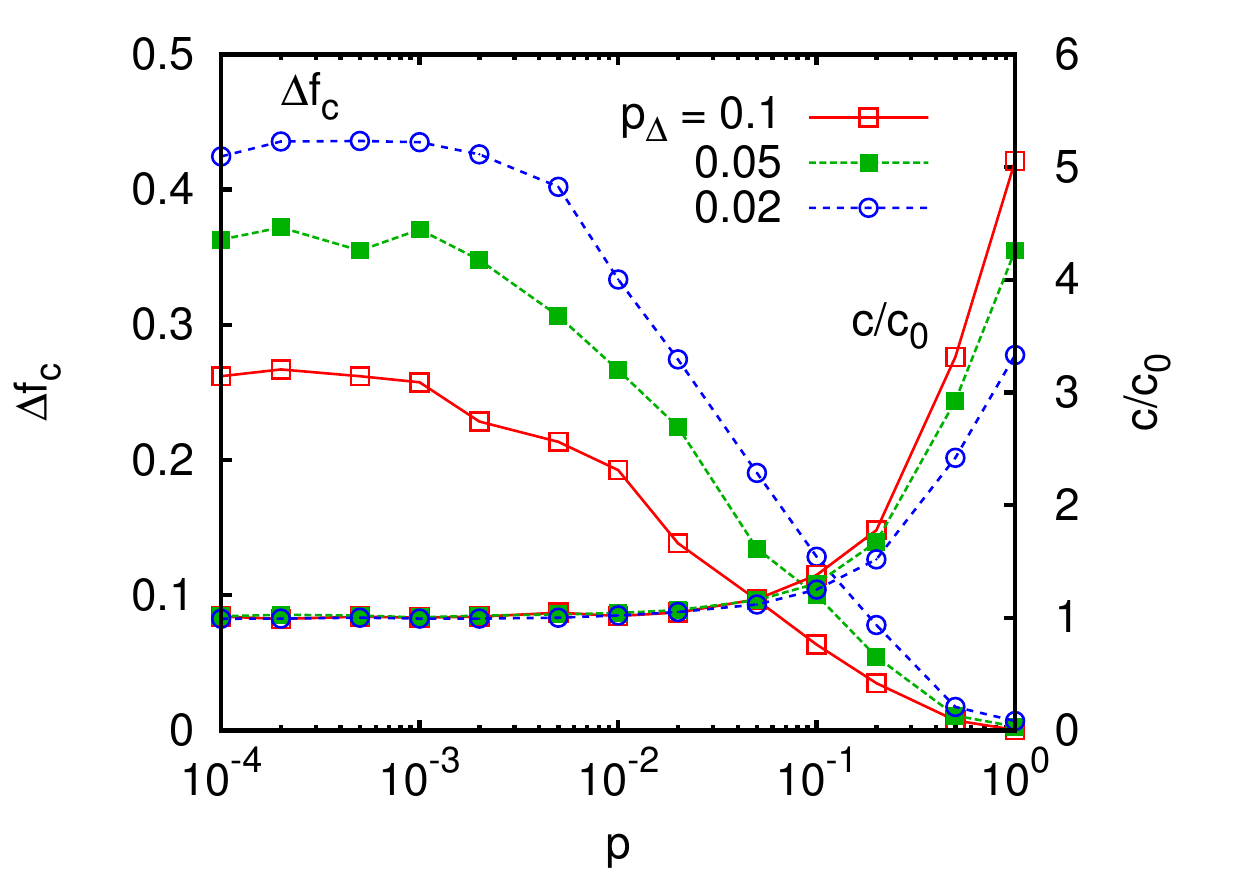}
}
\subfigure{
\includegraphics[width=.47\textwidth]{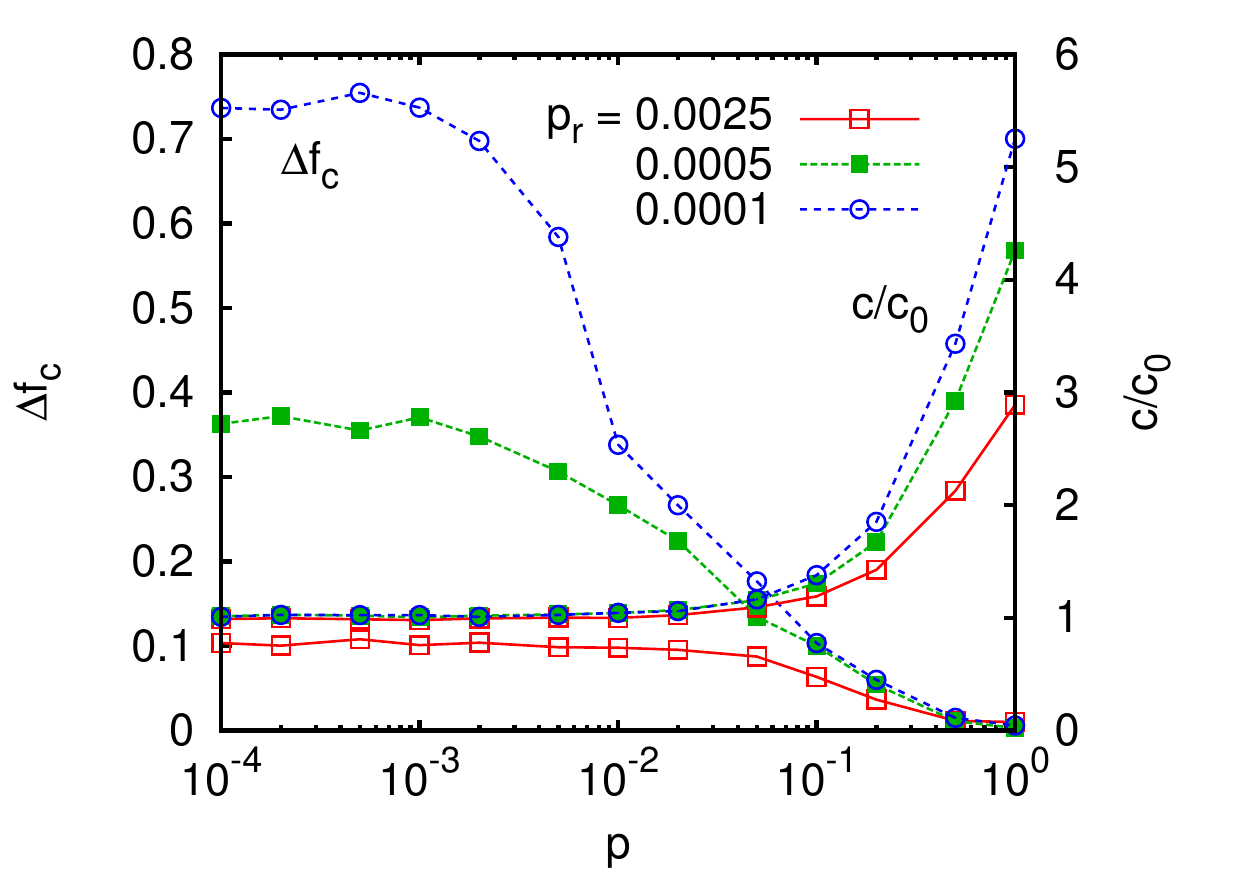}
}
\subfigure{
\includegraphics[width=.47\textwidth]{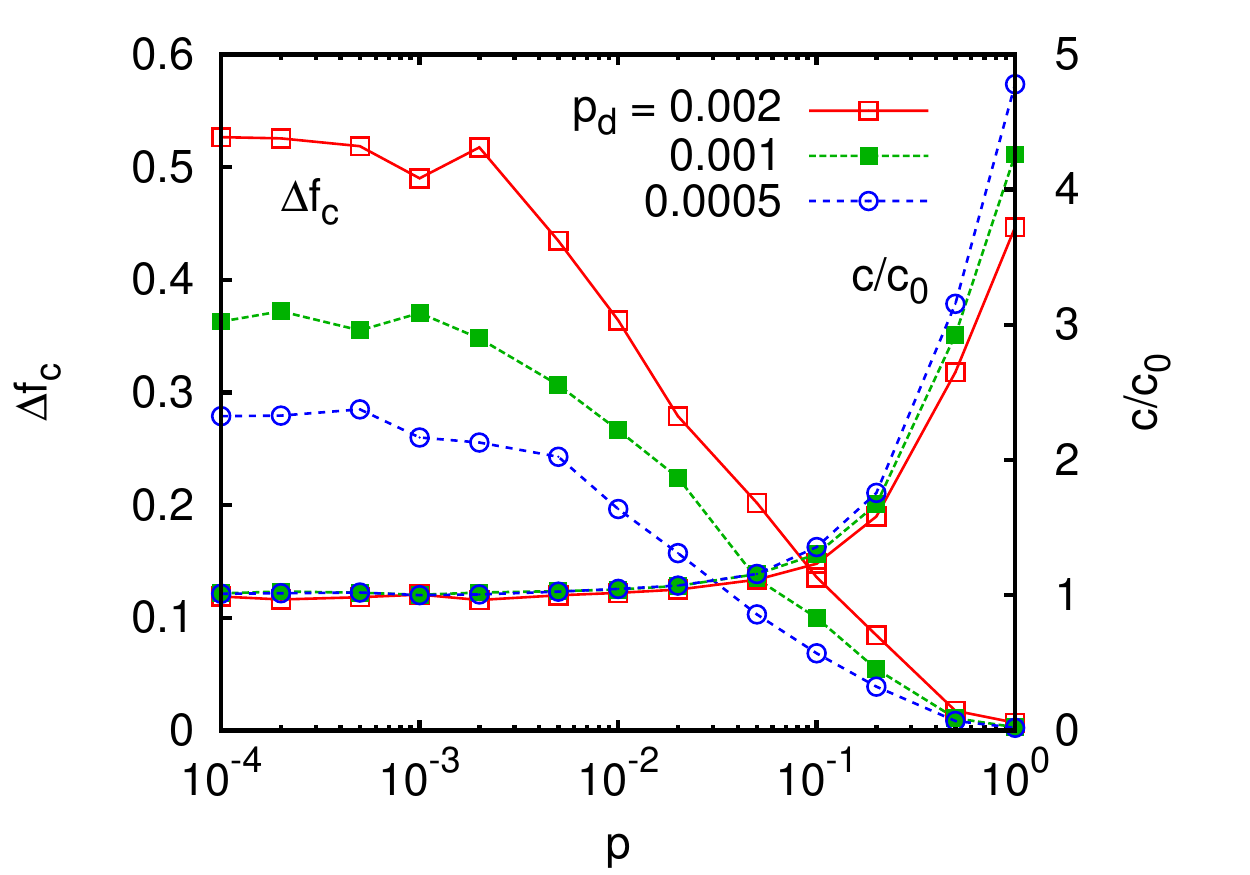}
}
\caption{
(Color online)
These figures show the same quantities as Fig.~\ref{fig:df_cc_cs} for several values of $p_\Delta$ (a), $p_r$ (b), and $p_d$ (c).
The results are obtained by simulation of $N=10000$ and averaged over $20$ independent samples.
}
\label{fig:trial_with_different_parameters}
\end{center}
\end{figure}

We tested other parameters for copy-and-shuffle model in order to verify the results are robust against the change of parameters.
Figure~\ref{fig:trial_with_different_parameters} shows the results when the parameters $p_\Delta$, $p_r$, and $p_d$ are modified from the ones used in Section.~\ref{subsec:model_multi_layer}.
All the tested results are qualitatively similar to Fig.~\ref{fig:df_cc_cs}: $\Delta f_c$ decreases to zero more quickly than the increase in $\overline{c}/\overline{c}_0$ when $p$ is increased.

\bibliography{complex-network}

\end{document}